\documentclass[a4paper,11pt]{article}

\usepackage[top=1in, bottom=1in, left=1in, right=1in]{geometry}

\usepackage{amsmath}
\usepackage{bm}
\usepackage{graphicx}
\usepackage[caption=false, justification=centering]{subfig}
\usepackage{url}
\usepackage{hyperref}

\usepackage{authblk}
\usepackage{datetime}

\begin{document}

%
%


\title{Inversion of Multi-frequency Data with the Cross-Correlated Contrast Source Inversion Method}

%
%




\author[1]{Shilong Sun\thanks{ShilongSun@icloud.com}} 
\author[1]{Bert Jan Kooij\thanks{B.J.Kooij@tudelft.nl}}
\author[1]{Alexander G. Yarovoy\thanks{A.Yarovoy@tudelft.nl}}

\affil[1]{Department of Microelectronics, Delft University of Technology, Mekelweg 4, 2628 CD, Delft, The Netherlands}

\renewcommand\Authands{ and }

\newdate{date}{4}{4}{2018}
\date{\displaydate{date}}

\maketitle

\begin{abstract}

	Cross-correlated contrast source inversion (CC-CSI) is a non-linear iterative inversion method that is proposed recently for solving the inverse scattering problems. In CC-CSI, a cross-correlated error is constructed and introduced to the cost functional, which improves the inversion ability when compared to the classical design of the cost functional by exploiting the mismatch between the data error and state error. In this paper, the multi-frequency inversion for electromagnetic waves is considered and a multi-frequency version of CC-CSI is proposed. Numerical and experimental inversion results of both transverse magnetic (TM) and transverse electric (TE) polarization demonstrate that, when multi-frequency data are available, CC-CSI still outperforms the multiplicative-regularized CSI method (MR-CSI) in the inversion of more complicated scatterers.

\end{abstract}

\section{Introduction}

    Determining the inhomogeneity in a certain region of medium by probing the scattered fields (electromagnetic and acoustic) is a common problem arising from many different fields in science and engineering such as remote sensing, biomedical imaging, geophysical exploration, nondestructive testing, etc. The information of interest could be the morphological information or the values of the inhomogeneities. In this paper, efforts have been made to study the inversion of the dielectric parameters, which is referred to as the inverse medium problem for electromagnetic waves \cite{colton2013inverse}. The research output is of course applicable for achieving the support of the objects, and it also applies to the inversion for acoustic waves.

    Since the inverse scattering problem we will be discussing is in the resonance range, i.e., the wavelength is comparable to the dimension of the objects, it turns out to be inherently nonlinear. Severe ill-posedness also accompanies this inverse scattering problem, which comes from both the inherent challenges of the physical problem itself and the incompleteness and/or inaccuracy of the measurement data domain. There are at least two main macro-classes of fully-nonlinear iterative inversion approaches: stochastic methods and deterministic ones. In cases where the dimension of the solution space is not huge, stochastic global optimization methods \cite{caorsi2000ACom,rocca2009evolutionary,rocca2011Diff,salucci2017multifrequency} are good candidates to search for the global optimal solution. Among the deterministic approaches, one is to use the linear sampling method together with a knowledge of the first transmission eigenvalue \cite{cakoni2011linear} or with several so-called ``virtual experiments'' \cite{crocco2012linear}. A more accurate alternative is to recover the dielectric parameters and the total fields iteratively using an optimization method \cite{wang1989iterative,kleinman1992modified,kleinman1993extended,van1997contrast,di2009numerical}. The iterative algorithms have been further improved over the recent decades by using regularization constraint \cite{van1999extended,bauer2009iteratively,sun2017Linearied}, multi-scaling technique \cite{caorsi2003new}, wavelet transformation \cite{li2013contrast}, etc. As an example, sparsity constraints can be appropriately exploited and compressive sensing-based techniques are proved to be effective in both qualitative imaging \cite{Gurbuz2009ACom,sun2017ALinearModel,sun2018ALinear} and quantitative ones \cite{oliveri2011bayesian,Poli2013MT,Ambrosanio2015ACom,sun2017Linearied}. However, it is worth noting that the cost functional remains unchanged in the aforementioned research work, which consists of two error terms: the data error and the state error. This classical construction of the cost functional has been changed recently by the work of \cite{sun2017Cross}, in which a cross-correlated error was proposed as a measure of the mismatch between the two error terms. Subsequently, the cross-correlated error was introduced into the cost function, which leads to a novel inversion method, referred to as the cross-correlated contrast source inversion (CC-CSI) method. It has been demonstrated that CC-CSI shows better inversion performance than the classical contrast source inversion (CSI) method \cite{van1997contrast} and multiplicative regularized CSI (MR-CSI) method \cite{van1999extended}, especially for inverting more complicated objects of higher contrast values. However, CC-CSI was only tested with single frequency in the work of \cite{sun2017Cross}, and it is still open whether it shows superiority when multi-frequency data is available, because the reliability of MR-CSI can be improved indeed by exploitation of multiple frequencies \cite{bloemenkamp2001inversion}.

    In this paper, we consider the inversion of multi-frequency electromagnetic data with CC-CSI. A multi-frequency version of CC-CSI (MF-CC-CSI) is proposed, which is able to process the multi-frequency data simultaneously. To validate the advantage of the proposed method, comparison has been made to the multi-frequency version of MR-CSI (MF-MR-CSI) proposed in \cite{bloemenkamp2001inversion}. Numerical results show that, compared to MF-MR-CSI, a more stabilized solution can be obtained by MF-CC-CSI. The $2$-nd Fresnel datasets in the year of 2005 \cite{geffrin2005free} have been selected as the experimental data for further validation. In the remainder of this paper, both transverse magnetic (TM) and transverse electric (TE) polarizations are considered. Problem statement and formulation of the proposed MF-CC-CSI method are introduced in Section \ref{sec.formulation}, the numerical simulation is presented in Section \ref{sec.numsim}, and the experimental data inversion is given in Section \ref{sec.exp}. The main body of this paper is finalized with conclusion in Section \ref{sec.conclusion}.

\section{Problem Statement and Formulation}\label{sec.formulation}

\subsection{Problem Statement}

    The multi-frequency inversion problem discussed in this paper is assumed to be non-dispersive, i.e., the contrast to be inverted is independent of frequency. Let us consider the canonical 2-D inverse scattering problem in a known background $\mathcal{D}$. The measurement domain $\mathcal{S}$ contains the sources denoted by the subscript $p\in\{1,2,3...,P\}$ and the receivers denoted by the subscript $q\in\{1,2,3,...,Q\}$. Equal subscript means the same position. We use a right-handed coordinate system in which the unit vector in the invariant direction points out of the paper. The time factor of $\exp(\text{i}\omega t)$ is considered in this paper, where $\text{i}^2=-1$.

    Now the data equation and the state equation can be formulated based on finite difference frequency domain (FDFD) scheme as (see \cite{sun2017Cross})
    \begin{subequations}
        \begin{align}
            \bm{y}_p &= \bm{\Phi}_{p}\omega^2 \bm{j}_{p}, \quad \bm{x}\in\mathcal{S},\\
            \bm{j}_{p} &= \bm{\chi}\bm{e}^{\text{inc}}_{p}+\bm{\chi}\bm{A}^{-1}\omega^2\bm{j}_{p}, \quad \bm{x}\in\mathcal{D},
        \end{align}
    \end{subequations}
    respectively, with $p = 1,2,\dots,P$, where, $\bm{y}_p$ represents the measurement data; $\bm{\Phi}_{p}:=\mathcal{M}_{\mathcal{S},p} \bm{A}^{-1}$ is the measurement matrix; $\mathcal{M}_{\mathcal{S},p}$ is an operator that interpolates field values defined at the finite-difference grid points to the appropriate receiver positions; $\bm{A}$ is the FDFD stiffness matrix; $\bm{e}^{\text{inc}}_{p}$ and $\bm{e}_p$ are the incident and total electric fields in the form of a column vector; $\bm{\chi}$ is the complex contrast consisting of the contrast permittivity, $\Delta\bm{\varepsilon}$, and the contrast conductivity, $\Delta\bm{\sigma}$, i.e., $\bm{\chi}:=\Delta\bm{\varepsilon}-\text{i}\Delta\bm{\sigma}/\omega$; $\bm{j}_{p}:=\bm{\chi} \bm{e}_p$ is defined as the component-wise multiplication of the contrast and the total fields, which is referred to as the contrast sources. In the remainder of this paper, $\omega^2$ is incorporated into $\bm{\Phi}_{p}$ and $\bm{A}$ for the sake of conciseness. Both of the matrices contain the background information. The inverse problem is to reconstruct the contrast $\bm{\chi}$ from the incomplete and/or inaccurate measurement data, $\bm{y}_p$.

\subsection{Multi-frequency CC-CSI}

\subsubsection{Modified Cost Functional, \texorpdfstring{$\mathcal{C}_{\text{MF-CC-CSI},\ell-1/2}$}{Lg}}

    Consider the inversion of multi-frequency data, the subscript $i$ is used to represent the $i$-th frequency. The data error equation and the state error equation for the update of the contrast sources are defined respectively as follows
    \begin{subequations}
        \begin{align}
            \bm{\rho}_{p,i,\ell-1/2} &= \bm{y}_{p,i}-\bm{\Phi}_{p,i}\bm{j}_{p,i,\ell-1}, \label{eq.DataStateEq.data}\\
            \bm{\gamma}_{p,i,\ell-1/2} &= \bm{\chi}_{i,\ell-1}\bm{e}^{\text{inc}}_{p,i}+\bm{\chi}_{i,\ell-1}\bm{A}_i^{-1}\bm{j}_{p,i,\ell-1}-\bm{j}_{p,i,\ell-1}.\label{eq.DataStateEq.state}
        \end{align}
    \end{subequations}
    with $p=1,2,3,\cdots,P$, $i= 1,2,3,\cdots,I$. Here, $(\ell-1/2)$ means the update of the contrast sources taking place after the $(\ell-1)$-th iteration and before the $\ell$-th update of the contrast. Equation~\eqref{eq.DataStateEq.data} is in the measurement domain and Equation~\eqref{eq.DataStateEq.state} is in the field domain. The latter is used to monitor the behavior of the solution in the field domain and check if it satisfies Maxwell's equations. Since Equation~\eqref{eq.DataStateEq.state} is always not perfectly satisfied, the ``$=$'' is supposed to be a ``$\approx$''. Note that the solution is monitored only in the field domain, the mismatch in the field domain should also be monitored back in the measurement domain. Otherwise, the design of the cost functional is logically not complete. In order to fill this gap, we define a new equation in the measurement domain as follows
    \begin{equation}
        \bm{\xi}_{p,i,\ell-1/2} = \bm{y}_{p,i} - \bm{\Phi}_{p,i}\left(\bm{\chi}_{i,\ell-1}\bm{e}^{\text{inc}}_{p,i}+\bm{\chi}_{i,\ell-1}\bm{A}_i^{-1}\bm{j}_{p,i,\ell-1}\right),
    \end{equation}
    with $p=1,2,3,\cdots,P$, $i= 1,2,3,\cdots,I$. We refer to this equation as multi-frequency cross-correlated error equation. Consequently, the cost functional for the update of the contrast sources is defined as follows
    \begin{equation}
        \begin{split}
            \mathcal{C}_{\text{MF-CC-CSI},\ell-1/2} = &\sum_{i=1}^I\eta_i^{\mathcal{S}}\sum_{p=1}^P\left\|\bm{\rho}_{p,i,\ell-1/2}\right\|_{\mathcal{S}}^2 +
            \sum_{i=1}^I\eta_{i,\ell-1}^{\mathcal{D}}\sum_{p=1}^P\left\|\bm{\gamma}_{p,i,\ell-1/2}\right\|_{\mathcal{D}}^2 + \\
            &\sum_{i=1}^I\eta_i^{\mathcal{S}}\sum_{p=1}^P\left\|\bm{\xi}_{p,i,\ell-1/2}\right\|_{\mathcal{S}}^2,
        \end{split}
    \end{equation}
    where, $\eta_i^{\mathcal{S}}$ and $\eta_{i,\ell-1}^{\mathcal{D}}$ are defined as
    \[
        \eta_i^{\mathcal{S}} = \sum_{p=1}^P\left\|\bm{y}_{p,i}\right\|_{\mathcal{S}}^2,\ \text{and}\ \ \eta_{i,\ell-1}^{\mathcal{D}} = \sum_{p=1}^P\left\|\bm{\chi}_{i,\ell-1}\bm{e}^{\text{inc}}_{p,i}\right\|_{\mathcal{D}}^2,
    \]
    respectively.

\subsubsection{Updating the Contrast Sources}

    The gradient (Fr{\'e}chet derivative) of the modified cost functional with respect to the contrast source $\bm{j}_{p,i}$ is
    \begin{equation}
        \begin{split}
        \bm{g}_{p,i} = & -2\eta_i^\mathcal{S}\bm{\Phi}_{p,i}^H\bm{\rho}_{p,i,\ell-1}+2\eta_{i,\ell-1}^\mathcal{D}\left(\bm{\chi}_{i,\ell-1}\bm{A}_i^{-1}-\bm{I}\right)^H\bm{\gamma}_{p,i,\ell-1}- \\
            & 2\eta_i^\mathcal{S}\left(\bm{\Phi}_{p,i}\bm{\chi}_{i,\ell-1}\bm{A}_i^{-1}\right)^H\bm{\xi}_{p,i,\ell-1}.
        \end{split}
    \end{equation}
    Now suppose $\bm{j}_{p,i,\ell-1}$ and $\bm{\chi}_{i,\ell-1}$ are known, then we update $\bm{j}_{p,i,\ell-1}$ by
    \begin{equation}
        \bm{j}_{p,i,\ell} = \bm{j}_{p,i,\ell-1} + \alpha_{p,i,\ell}\bm{\nu}_{p,i,\ell},
    \end{equation}
    where $\alpha_{p,i,\ell}$ is constant and the update direction $\bm{\nu}_{p,i,\ell}$ is chosen as the Polak-Ribi{\`e}re conjugate gradient directions, which is given by
    \begin{equation}
        \bm{\nu}_{p,i,\ell} =
        \begin{cases}
            0, &\quad \ell=0, \\
            \bm{g}_{p,i,\ell}+\frac
                 {
                   \sum_{p'}
                   \left\langle
                     \bm{g}_{p',i,\ell},\bm{g}_{p',i,\ell}-\bm{g}_{p',i,\ell-1}
                   \right\rangle
                   _{\mathcal{D}}
                 }
                 {
                   \sum_{p'}
                     \left\|
                       \bm{g}_{p',i,\ell-1}
                     \right\|
                     _{\mathcal{D}}^2
                 }
                 \bm{\nu}_{p,i,\ell-1}, &\quad \ell\geq1.
        \end{cases}
    \end{equation}
    where
    \begin{equation}
        \bm{g}_{p,i,\ell} = \left.\bm{g}_{p,i}\right|_{\bm{j}_{p,i}=\bm{j}_{p,i,\ell-1}}.
    \end{equation}
    The step size $\alpha_{p,i,\ell}$ is the minimizer of the cost functional
    \begin{equation}
        \left.\mathcal{C}_{\text{MF-CC-CSI},\ell-1/2}\right|_{\bm{j}_{p,i}=\bm{j}_{p,i,\ell-1} + \alpha_{p,i}\bm{\nu}_{p,i,\ell}}.
    \end{equation}
    See Appendix \ref{sec.appendix-1-2} for the derivation of $\alpha_{p,i,\ell}$. Following the update of the contrast sources, the total fields are updated by
    \begin{equation}
        \bm{e}^{\text{tot}}_{p,i,\ell} = \bm{e}^{\text{tot}}_{p,i,\ell-1} + \alpha_{p,i,\ell}\bm{A}_i^{-1}\bm{\nu}_{p,i,\ell}.
    \end{equation}
    Consequently, the state error and the cross-correlated error are supposed to be updated as well: $\bm{\gamma}_{p,i,\ell-1} \rightarrow \bm{\gamma}_{p,i,\ell}$, $\bm{\xi}_{p,i,\ell-1} \rightarrow \bm{\xi}_{p,i,\ell}$.

\subsubsection{Modified Cost Functional, \texorpdfstring{$\mathcal{C}_{\text{MF-CC-CSI},\ell}$}{Lg}, and Updating the Contrast}

    Suppose $\bm{\chi}_{i,\ell-1}$ is known and consider the relation
    \begin{equation}
        \bm{\chi}_{1,\ell-1} = \Re\left\{\bm{\chi}_{i,\ell-1}\right\} + \frac{\omega_1}{\omega_i}\Im\left\{\bm{\chi}_{i,\ell-1}\right\}, \quad i = 1,2,3,\cdots,I,
    \end{equation}
    where, $\Re$ and $\Im$ respectively represent the real parts and imaginary parts of complex numbers, we therefore define $\bm{\chi}_{\ell} =\bm{\chi}_{1,\ell}$. Once the contrast source $\bm{j}_{p,i,\ell}$ is determined, we update the contrast by
    \begin{equation}\label{eq.chiupdatingMF}
        \bm{\chi}_{\ell} = \bm{\chi}_{\ell-1} + \beta_{\ell}\bm{\nu}^{\chi}_{\ell},
    \end{equation}
    where, $\beta_\ell$ is the step size, and the update direction, $\bm{\nu}^{\chi}_{\ell}$, is chosen to be the Polak-Ribi{\`e}re conjugate gradient directions, which is given by
    \begin{equation}\label{eq.chinuMF}
        \bm{\nu}^{\chi}_{\ell} =
        \begin{cases}
        0, &n=0,\\
        \bm{g}^{\chi}_{\ell}+\frac
        {
        \left\langle
         \bm{g}^{\chi}_{\ell},\bm{g}^{\chi}_{\ell}-\bm{g}^{\chi}_{\ell-1}
        \right\rangle
        _{\mathcal{D}}
        }
        {
         \left\|
           \bm{g}^{\chi}_{\ell-1}
         \right\|
         _{\mathcal{D}}^2
        }
        \bm{\nu}^{\chi}_{\ell-1}, \quad &n\geq1,
        \end{cases}
    \end{equation}
    where, $\bm{g}^{\chi}_{\ell}$ is the preconditioned gradient of the modified multi-frequency cost functional for updating the contrast
    \begin{equation}
        \mathcal{C}_{\text{MF-CC-CSI},\ell} =
        \sum_{i=1}^I\eta_{i,\ell-1}^{\mathcal{D}}\sum_{p=1}^P\left\|\bm{\gamma}_{p,i,\ell}\right\|_{\mathcal{D}}^2 + \sum_{i=1}^I\eta_i^{\mathcal{S}}\sum_{p=1}^P\left\|\bm{\xi}_{p,i,\ell}\right\|_{\mathcal{S}}^2.
    \end{equation}
    Here,
    \begin{equation}
        \bm{\gamma}_{p,i,\ell} = \bm{\chi}_{i,\ell-1}\bm{e}^{\text{inc}}_{p,i}+\bm{\chi}_{i,\ell-1}\bm{A}_i^{-1}\bm{j}_{p,i,\ell}-\bm{j}_{p,i,\ell},
    \end{equation}
    \begin{equation}
        \bm{\xi}_{p,i,\ell} = \bm{y}_{p,i} - \bm{\Phi}_{p,i}\left(\bm{\chi}_{i,\ell-1}\bm{e}^{\text{inc}}_{p,i}+\bm{\chi}_{i,\ell-1}\bm{A}_i^{-1}\bm{j}_{p,i,\ell}\right).
    \end{equation}
    Specifically, $\bm{g}^{\chi}_{\ell}$ is given by
    \begin{equation}
        \bm{g}^{\chi}_{\ell} =  \frac{2\Re\left\{\sum_{i=1}^I\bm{g}^{\chi}_{i,\ell}\right\}}
        {\sum_{i=1}^I\sum_{p=1}^P\bm{e}^{\text{tot}}_{p,i,\ell}\overline{\bm{e}^{\text{tot}}_{p,i,\ell}}}+\text{i}
        \frac{2\Im\left\{\sum_{i=1}^I\frac{\omega_1}{\omega_i}\bm{g}^{\chi}_{i,\ell}\right\}}
        {\sum_{i=1}^I\left(\frac{\omega_1}{\omega_i}\right)^2\sum_{p=1}^P\bm{e}^{\text{tot}}_{p,i,\ell}\overline{\bm{e}^{\text{tot}}_{p,i,\ell}}},
    \end{equation}
    where,
    \begin{equation}
        \bm{g}^{\chi}_{i,\ell} = \eta^{\mathcal{D}}_{i,\ell-1}\sum_{p=1}^P\overline{\bm{e}^{\text{tot}}_{p,i,\ell}}\bm{\gamma}_{p,i,\ell}-\eta_i^{\mathcal{S}}\sum_{p=1}^P\overline{\bm{e}^{\text{tot}}_{p,i,\ell}}\bm{\Phi}_{p,i}^H\bm{\xi}_{p,i,\ell}.
    \end{equation}
    The step size $\beta_\ell$ is determined by minimizing the updated cost function in the formulation of
    \begin{equation}
        \sum_{i=1}^I\frac{\sum_{p=1}^P\left\|\left(\bm\chi_{i,\ell-1}+\beta\bm\chi_{i,\ell}\right)\bm{e}^{\text{tot}}_{p,i,\ell}-\bm{j}_{p,i,\ell}\right\|_\mathcal{D}^2}
        {\sum_{p=1}^P\left\|\left(\bm\chi_{i,\ell-1}+\beta\bm\chi_{i,\ell}\right)\bm{e}_{p,i}^{\text{inc}}\right\|_\mathcal{D}^2} +
        \sum_{i=1}^I\eta^\mathcal{S}_i\sum_{p=1}^P\left\|\bm{y}_{p,i}-\bm{\Phi}_{p,i}\left(\bm\chi_{i,\ell-1}+\beta\bm\chi_{i,\ell}\right)\bm{e}^{\text{tot}}_{p,i,\ell}\right\|_\mathcal{S}^2.
    \end{equation}
    This is a problem of finding the minimum of a single-variable function, which can be solved efficiently by the Brent's method \cite{brent1973algorithms,Forsythe1976computer}. It is worth noting that the objects are assumed to be isotropic in this paper. Namely, we assume $\bm\chi_{x}=\bm\chi_{y}$ when TE-polarized data is processed.

\subsubsection{Initialization}

    If no \textit{a priori} information about the objects is available, the contrast source is initialized using the values obtained by back-propagation \cite{van1997contrast}
    \begin{equation}\label{eq.initvjMF}
        \bm{j}_{p,i,0} = \frac{\left\|\bm{\Phi}_{p,i}^H\bm{y}_{p,i}\right\|^2_{\mathcal{D}}}{\left\|\bm{\Phi}_{p,i}\bm{\Phi}_{p,i}^H\bm{y}_{p,i}\right\|^2_{\mathcal{S}}}\bm{\Phi}_{p,i}^H\bm{y}_{p,i},
    \end{equation}
    and the starting value of the total field is
    \begin{equation}
        \bm{e}^{\text{tot}}_{p,i,0} = \bm{e}^{\text{inc}}_{p,i} + \bm{A}_i^{-1}\bm{j}_{p,i,0}.
    \end{equation}
    The contrast is initialized by (see \cite{bloemenkamp2001inversion})
    \begin{equation}\label{eq.initchiMF}
        \bm{\chi}_{1,0} = \frac{\Re\left\{\displaystyle\sum_{i=1}^I\sum_{p=1}^P\bm{j}_{p,i,0}\overline{\bm{e}^{\text{tot}}_{p,i,0}}\right\}}
        {\displaystyle\sum_{i=1}^I\sum_{p=1}^P\bm{e}^{\text{tot}}_{p,i,0}\overline{\bm{e}^{\text{tot}}_{p,i,0}}} + \text{i}
        \frac{\Im\left\{\displaystyle\sum_{i=1}^I\frac{\omega_1}{\omega_i}\sum_{p=1}^P\bm{j}_{p,i,0}\overline{\bm{e}^{\text{tot}}_{p,i,0}}\right\}}
        {\displaystyle\sum_{i=1}^I\left(\frac{\omega_1}{\omega_i}\right)^2\sum_{p=1}^P\bm{e}^{\text{tot}}_{p,i,0}\overline{\bm{e}^{\text{tot}}_{p,i,0}}}.
    \end{equation}

    Since free space is considered in the following examples, the contrast must have non-negative real part and non-positive imaginary part. Such properties are enforced in the following examples by simply setting the negative real part and the positive imaginary part of the contrast to zero following each update of the contrast for both MF-CC-CSI and MF-MR-CSI. In addition, the contrast sources and the contrast are initialized by Equation \eqref{eq.initvjMF} and Equation \eqref{eq.initchiMF}, respectively, for fair comparison.

\section{Numerical Simulation}\label{sec.numsim}

    In this section, both MF-CC-CSI and MF-MR-CSI are tested for comparison with a 2-D benchmark problem -- the ``Austria'' profile, which was also used in \cite{litman1998reconstruction,van2001contrast,van2003multiplicative,sun2017Cross}. The objects consist of two disks and one ring. Let us first establish our coordinate system such that the $z$-axis is parallel to the axis of the objects. The disks of radius 0.2 m are centred at ($-0.3$, 0.6) m and ($0.3$, 0.6) m. The ring is centred at (0,$-0.2$) m, and it has an exterior radius of 0.6 m and an inner radius of 0.3 m. We assume that the three cylinders are made of same material. Two sets of dielectric parameters are considered, which are $\varepsilon_r=3$, $\sigma=5$ mS/m, and $\varepsilon_r=10$, $\sigma=10$ mS/m, respectively. Since we assume a free space environment, the contrast values are $\Delta\varepsilon_r=2$, $\Delta\sigma=5$ mS/m, and $\Delta\varepsilon_r=9$, $\Delta\sigma=10$ mS/m, respectively. The forward scattering problems are solved by a MATLAB-based ``MaxwellFDFD'' package \cite{W.Shin2013} with a fine and non-uniform grid size of $\lambda/(45\sqrt{\varepsilon_r})$.

\subsection{Measurement configuration}

    In order to approach the realistic situation, we selected a measurement configuration which is similar to the experiment conducted by the Remote Sensing and Microwave Experiments Team at the Institut Fresnel, France \cite{0266-5611-17-6-301}. An emitter is fixed on the circular rail, while a receiver is rotating with the arm around a vertical cylindrical target. The targets rotated from 0$^{\circ}$ to 350$^{\circ}$ in steps of 30$^{\circ}$, and the receiver rotated from 60$^{\circ}$ to 300$^{\circ}$ in steps of 5$^{\circ}$. The distance from the emitter/receiver to the origin is $3$ m. Simulation is sequentially done without and with the objects at five frequencies of 0.1 GHz, 0.2 GHz, 0.3 GHz, 0.4 GHz, and 0.5 GHz. Scattered field data is obtained by subtracting the incident field data, $\bm{y}^{\text{inc}}_p$, from the total field data, $\bm{y}^{\text{tot}}_p$.

\subsection{Inversion results}

    Although the incident fields in the inversion domain can be easily obtained in numerical simulations, it is not available in real applications. Therefore, we consider the modeling of the incident fields using the collected incident field data. We select the approach reported in \cite{bloemenkamp2001inversion}. Specifically, the transmitting antenna is approximated by line source parallel to the cylindrical objects. The incident field by the line source is calibrated by multiplying a complex ratio which is calculated using only the value of the field when the transmitting and the receiving antenna are in opposite directions. For each frequency and source position, one complex calibration factor is determined. To simulate the inevitable measurement error in real experiments, additive Gaussian random noise, $\bm{n}_p$, is added directly to the scattered field data. In addition, the noise is also added to the total field data and the incident field data by $\bm{y}^{\text{tot}}_p+\bm{n}_p/2$ and $\bm{y}^{\text{inc}}_p-\bm{n}_p/2$, respectively. In doing so, the disturbance of the noise is reflected not only in the scattered field data, but also in the modelling of the incident fields, which better matches the real situation. To appraise the inversion accuracy, let us define the reconstruction error as follows
    \begin{equation}\label{eq.err}
        err := \left.\left\|\hat{\bm{\chi}}-\bm{\chi}\right\|_2\middle /\left\|\bm{\chi}\right\|_2\right.,
    \end{equation}
    where, $\hat{\bm{\chi}}$ and $\bm{\chi}$ are the reconstructed contrast and the exact contrast, respectively. Since large values of frequency tolerates large errors of estimated conductivity (see Subsection \ref{subsec.InvResExp} for more discussion), $\hat{\bm{\chi}}$ and $\bm{\chi}$ in Equation \eqref{eq.err} correspond to the highest frequency to ensure that the imaginary part of the contrast error is not over amplified in the evaluation of the inversion accuracy. In the following two cases, the inversion domain is restricted to [$-1.20$, $1.20$] $\times$ [$-1.20$, $1.20$] m$^2$ and discretized into uniform grids with a size of $0.30\times0.30$ mm$^2$.

\subsubsection{Case 1: \texorpdfstring{$\Delta\varepsilon_r=2$, $\Delta\sigma=5$}{Lg} mS/m}

    \begin{figure}[!t]
        \centering
        \includegraphics[width=1.00\linewidth] {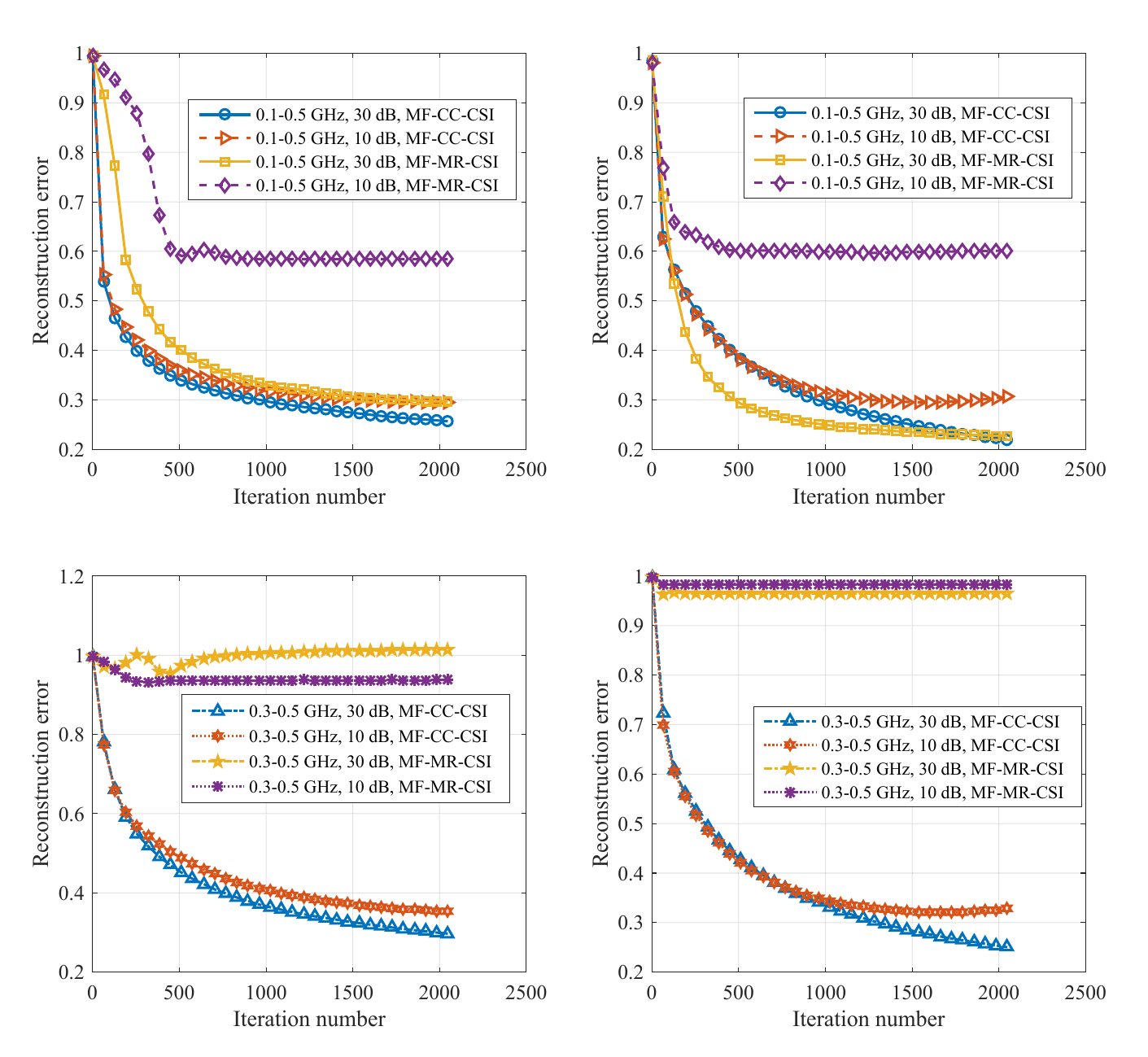}%
        \caption{Inversion error curves of MF-CC-CSI and MF-MR-CSI in terms of iteration number in case 1 ($\Delta\varepsilon_r=2$, $\Delta\sigma=5$ mS/m) by processing the TM-polarized (left) and TE-polarized (right) data. Different frequency bands of 0.1-0.5 GHz and 0.3-0.5 GHz and different SNRs of 30 dB and 10 dB are considered.}
        \label{fig:Sim1}
    \end{figure}

    Let us first consider the lower contrast case, i.e., $\Delta\varepsilon_r=2$, $\Delta\sigma=5$ mS/m. MF-CC-CSI and MF-MR-CSI are used to process the TM- and TE-polarized numerical data respectively with 2048 iterations. Different frequency bands of 0.1-0.5 GHz and 0.3-0.5 GHz and different SNRs of 30 dB and 10 dB are tested. Figure \ref{fig:Sim1} shows the reconstruction error curves in term of the iteration number, which indicates obviously that MF-CC-CSI obtains less reconstruction errors compared to MF-MR-CSI when same frequency band and SNR are used. It is also easy to observe that noise indeed leads to a degradation of the inversion accuracy (see the error curves of MF-CC-CSI shown in the right subfigures of Figure \ref{fig:Sim1} when processing the 10 dB SNR data). As the frequency goes up to 0.3-0.5 GHz, the reconstruction error of MF-MR-CSI remains close to 1, indicating that MF-MR-CSI completely fails to invert the data in this frequency band.

\subsubsection{Case 2: \texorpdfstring{$\Delta\varepsilon_r=9$, $\Delta\sigma=10$}{Lg} mS/m}

    \begin{figure}[!t]
        \centering
        \includegraphics[width=1.00\linewidth] {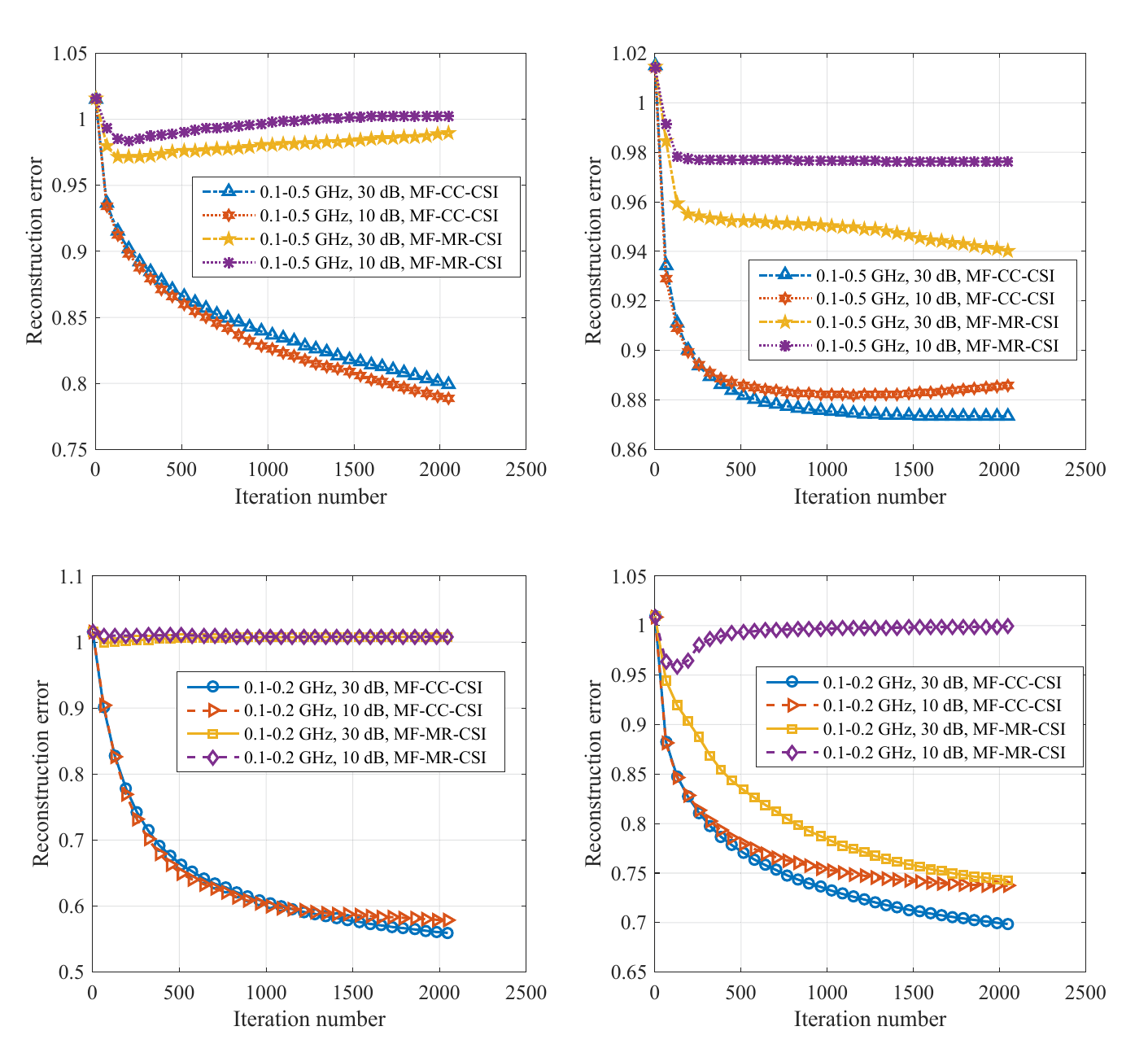}%
        \caption{Inversion error curves of MF-CC-CSI and MF-MR-CSI in terms of iteration number in case 2 ($\Delta\varepsilon_r=9$, $\Delta\sigma=10$ mS/m) by processing the TM-polarized (left) and TE-polarized (right) data. Different frequency bands of 0.1-0.5 GHz and 0.1-0.2 GHz and different SNRs of 30 dB and 10 dB are considered.}
        \label{fig:Sim2}
    \end{figure}

    In this case, the contrast increases to $\Delta\varepsilon_r=9$, $\Delta\sigma=10$ mS/m, which is supposed to be more challenging. We have also considered the frequency band 0.1-0.5 GHz, and from Figure \ref{fig:Sim2} we observe that both MF-CC-CSI and MF-MR-CSI obtained large reconstruction errors. However, the former obtained less and smoothly decreasing reconstruction errors than the latter. Now let us remove the high frequency data and consider the frequency band 0.1-0.2 GHz, one can see from the bottom two subfigures of Figure \ref{fig:Sim2} that the inversion accuracy of MF-CC-CSI has been improved; The reconstruction error curve of MF-MR-CSI started to decrease when processing the 0.1-0.2 GHz 30 dB synthetic data. Both of the two cases demonstrate that MF-CC-CSI is a more reliable inversion method in comparison to MF-MR-CSI.

\subsection{Analysis of global convergence capability}

\begin{figure}[!t]
        \centering
        \includegraphics[width=1.00\linewidth] {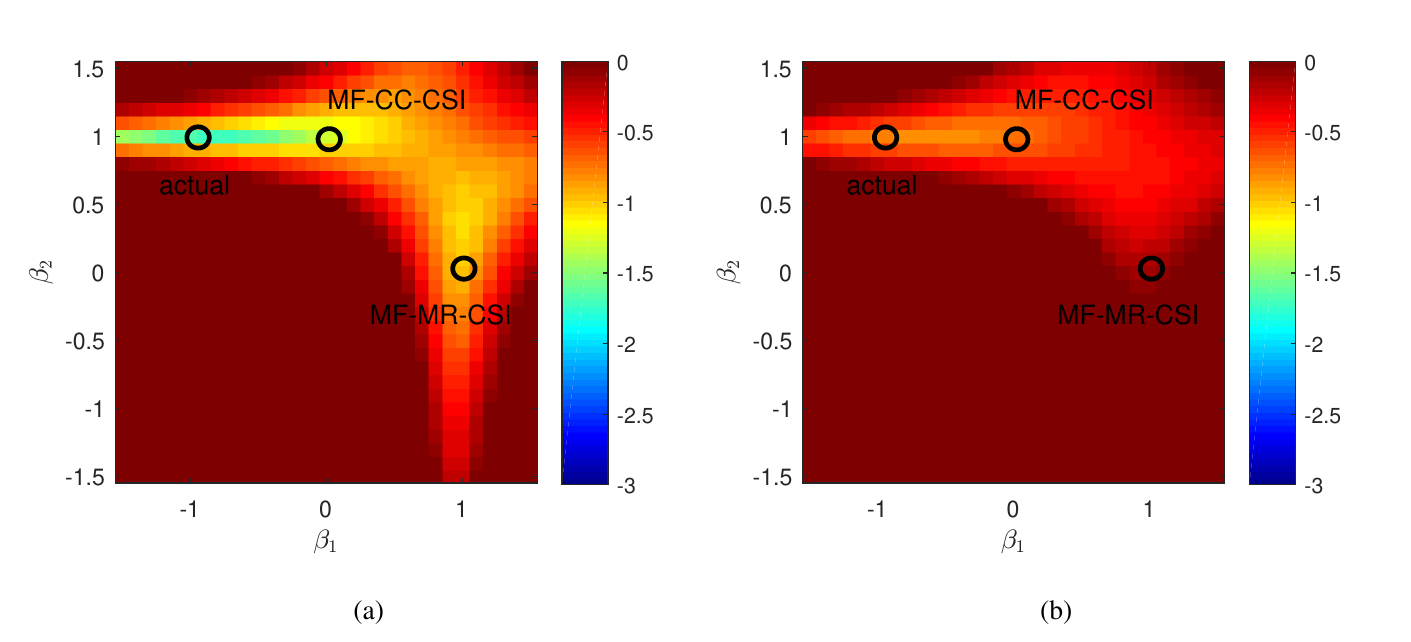}%
        \caption{Illustrative example: the 0.1-0.5 GHz TM example in Case 1. Behavior of the cost function, $\log_{10}\left\{\mathcal{C}_{\text{MF-CC-CSI}}\left(\beta_1, \beta_2\right)\right\}$, in the parametric range $-1.5\leq\beta_1\leq1.5$ and $-1.5\leq\beta_2\leq1.5$. (a) SNR$=30$ dB; (b) SNR$=10$ dB.}
        \label{fig:CostFun}
    \end{figure}

    In this subsection, let us have more insights on the enhanced global convergence capability of MF-CC-CSI by analyzing the behavior of the modified cost function in the solution space. Specifically, let us first define the solution as follows
    \begin{equation}
        \bm{x} := \left\{\bm{\chi}_i, \bm{e}^{\text{tot}}_{p,i}, \quad p=1,2,3,\cdots,P, \ i= 1,2,3,\cdots,I\right\}.
    \end{equation}
    The contrast sources are not considered here because, knowing the contrast and the total fields, it can be simply obtained. With two sampling parameters, $\beta_1$ and $\beta_2$, the solution space is defined as follows
    \begin{equation}
        \bm{x}\left(\beta_1, \beta_2\right) := \beta_2\left[\left(\beta_1+1\right)\bm{x}^{\text{opt}}_{\text{MF-CC-CSI}}-\beta_1\bm{x}^{\text{act}}\right]-\left(\beta_2-1\right)\beta_1\bm{x}^{\text{opt}}_{\text{MF-MR-CSI}},
    \end{equation}
    where, $\bm{x}^{\text{act}}$ is the actual solution, while $\bm{x}^{\text{opt}}_{\text{MF-CC-CSI}}$ and $\bm{x}^{\text{opt}}_{\text{MF-MR-CSI}}$ are the solutions of MF-CC-CSI and MR-MR-CSI, respectively. Now let us consider the 0.1-0.5 GHz TM example in Case 1 and plot the cost function value versus two solution-space sampling parameters, $\mathcal{C}_{\text{MF-CC-CSI}}\left(\beta_1, \beta_2\right)=\mathcal{C}_{\text{MF-CC-CSI}}\left\{\bm{x}\left(\beta_1, \beta_2\right)\right\}$, with $\beta_1, \beta_2\in[-1.5,1.5]$. Figure \ref{fig:CostFun} (a) and (b) show the behavior of the cost function, $\log_{10}\left\{\mathcal{C}_{\text{MF-CC-CSI}}\left(\beta_1, \beta_2\right)\right\}$, with different SNRs, 30 dB and 10 dB, respectively. From Figure \ref{fig:CostFun} we see that the MF-MR-CSI converges to a local minimum, while $\bm{x}^{\text{opt}}_{\text{MF-CC-CSI}}$ belongs to the attraction basin of the global optimum of the cost function, $\bm{x}^{\text{act}}$. In addition, Figure \ref{fig:CostFun} indicates that the modified cost function still shows the multiminima property and a good initial guess is still critical for preventing the occurrence of false solutions. By comparison of Figure \ref{fig:CostFun} (a) and Figure \ref{fig:CostFun} (b) we can also observe that the behavior of the cost function is affected by the noise level in the measurement data.

\begin{figure}[!t]
        \centering
        \includegraphics[width=1.00\linewidth] {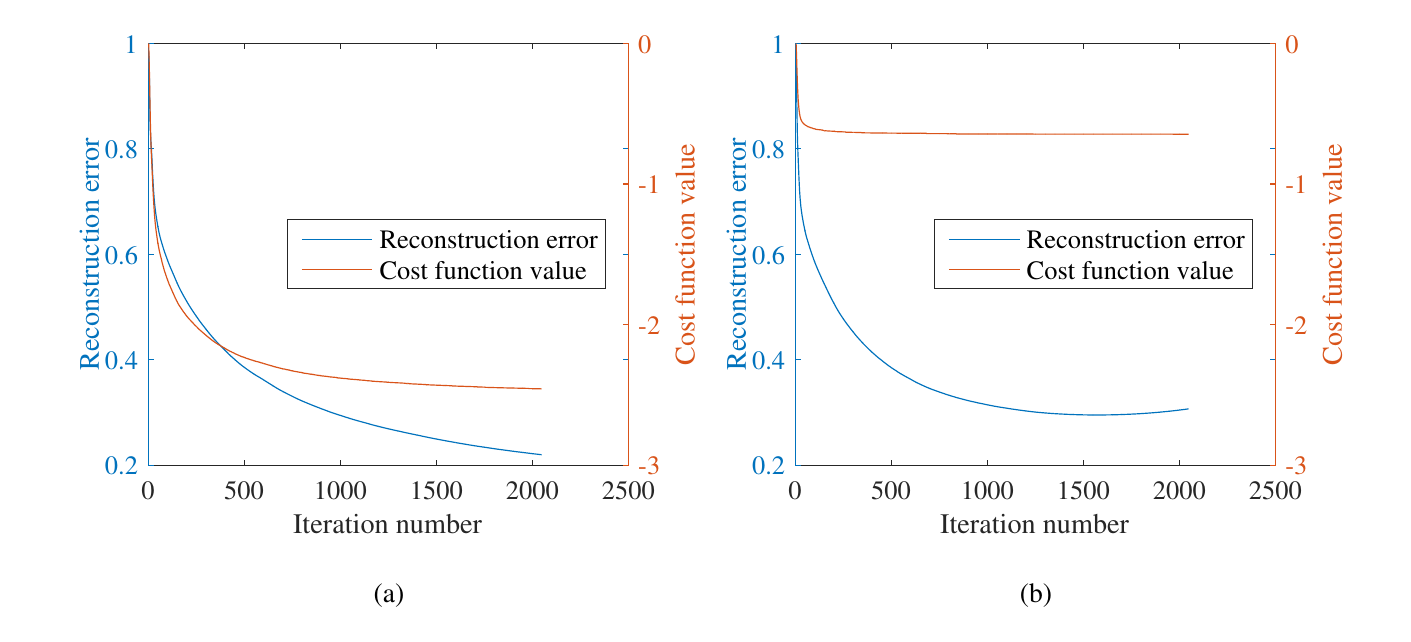}%
        \caption{Illustrative example: the 0.1-0.5 GHz TE example in Case 1. Behavior of the reconstruction error and the cost function ($\log_{10}\left\{\mathcal{C}_{\text{MF-CC-CSI}}\right\}$) of MF-CC-CSI. (a) SNR$=30$ dB; (b) SNR$=10$ dB.}
        \label{fig:ErrCostFun}
    \end{figure}

    Here we finally remark that a good termination condition is required for MF-CC-CSI. Otherwise, the inversion accuracy may deteriorate after a mount of iterations when the disturbance of noise to the measurement data is not negligible (see the top right figures in Figure \ref{fig:Sim1} and Figure \ref{fig:Sim2}). Now let us consider the 0.1-0.5 GHz TE example in Case 1, Figure \ref{fig:ErrCostFun} (a) and (b) show the behavior of the reconstruction error and the cost function ($\log_{10}\left\{\mathcal{C}_{\text{MF-CC-CSI}}\right\}$) of MF-CC-CSI with SNR$=30$ dB and 10 dB, respectively. One can observe that a threshold can hardly be determined to terminate the inversion algorithm without considering the noise level. In addition, the value of the contrast and the measurement configuration also affect the cost function curve and the convergence rate of the iterative inversion algorithms.

    One reasonable strategy in practice is to first estimate the noise level. Moreover, according to some \textit{a priori} information, it is possible to estimate the range of the contrast value. Based upon these information, one can do simulation using a typical benchmark inverse problem for a specific application, then obtain approximately how many iterations are required to get a reasonable inverted result for a specific application. Take the following experimental data inversion as an example. Considering the fact that the measurement configuration, the noise level and the range of the contrast value are similar to the numerical simulation, let us run 2048 iterations in the following experimental data inversion.

\section{Experimental Data}\label{sec.exp}

\subsection{Configuration}

    In this section, we consider the experiment carried out by Institut Fresnel in the year of 2005 \cite{geffrin2005free}. In this experiment, the receiver stays in the azimuthal plane (xoy) and rotates along two-thirds of a circle from 60$^\circ$ to 300$^\circ$ with the angular step being 1$^\circ$. The source antenna stays at the fixed location ($\theta = 0^\circ$). The targets are rotated to obtain different illumination incidences. The distance from the transmitter and receiver to the centre of the target domain is 1.67 m.

    To avoid the redundancy of our discussion, let us select the most complicated configuration that corresponds to the datasets \textit{FoamTwinDielTM} and \textit{FoamTwinDielTE}. The targets consist of one larger circular dielectric cylinder with a smaller one embedded inside and a smaller adjacent one outside (see Figure 5(c) in \cite{geffrin2005free}). Two smaller circular dielectric cylinders have relative permittivity values of $\varepsilon_r = 3\pm 0.3$ while the larger one has a relative permittivity value of $\varepsilon_r = 1.45\pm 0.15$. In this configuration, the targets are rotated from 0$^\circ$ to 340$^\circ$ with angular step of 20$^\circ$. $241\times 18$ measurements are obtained at each frequency (in TE polarization, only the component orthogonal to both the invariance axis of the cylinder and the direction of illumination is measured). The measurement configuration is given by Figure 1 of \cite{geffrin2005free}. To increase the inversion difficulty, let us assume the data in the low frequency band is not available anymore, and we only have the measurement data at 7 GHz, 8 GHz, 9 GHz, and 10 GHz.

\subsection{Inversion Results}\label{subsec.InvResExp}

    \begin{figure}[!t]
        \centering
        \includegraphics[width=1.00\linewidth] {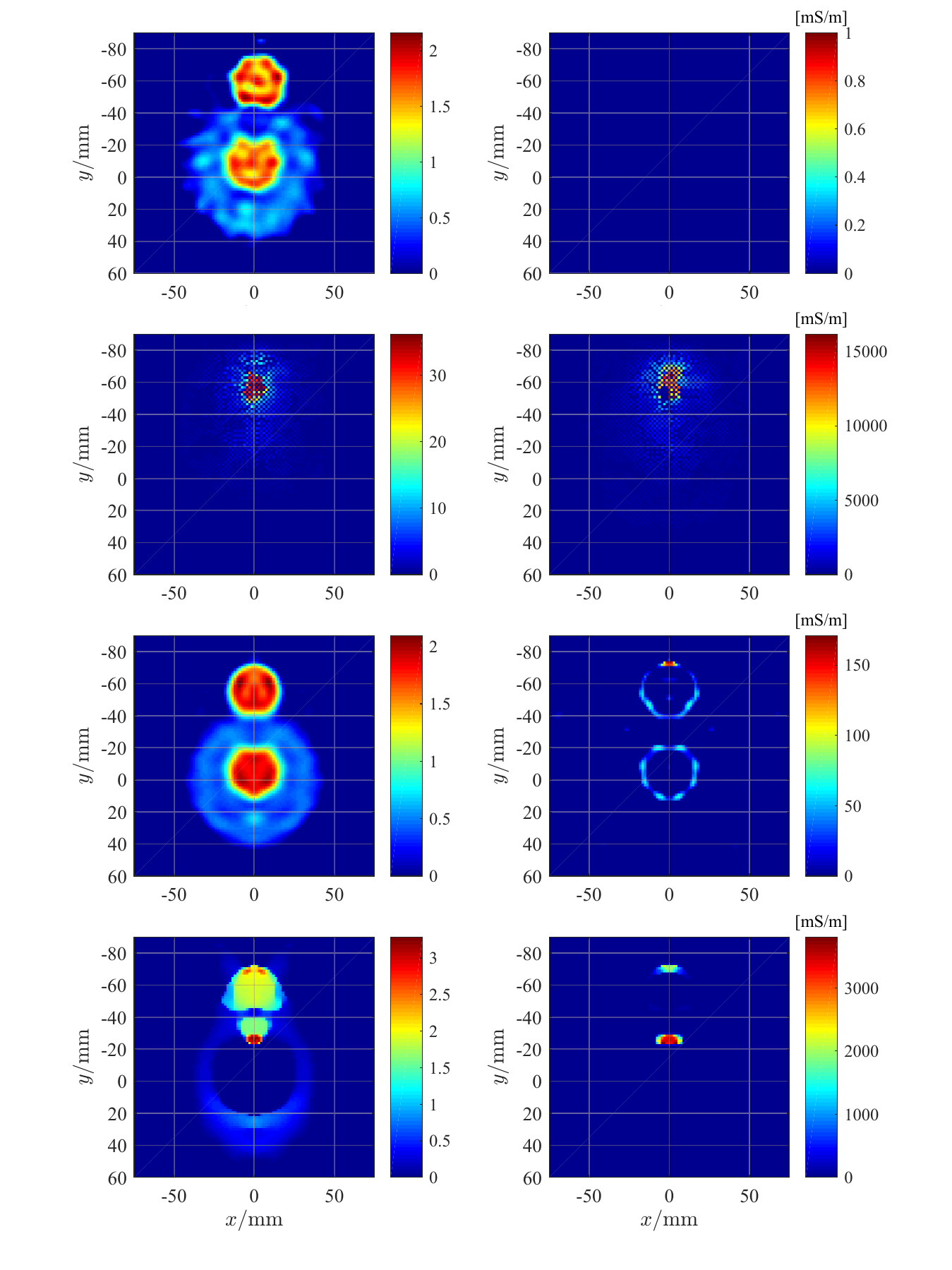}%
        \caption{Relative permittivity (left) and conductivity (right) of the inverted contrast by processing the multi-frequency dataset \textit{FoamTwinDielTM} (Line 1: MF-CC-CSI; Line 2: MF-MR-CSI) and its perfect synthetic data (Line 3: MF-CC-CSI; Line 4: MF-MR-CSI) at 7 GHz, 8 GHz, 9 GHz, and 10 GHz with 2048 iterations.}
        \label{fig:MFFoamTwinDielTMInv}
    \end{figure}

    \begin{figure}[!t]
        \centering
        \includegraphics[width=1.00\linewidth] {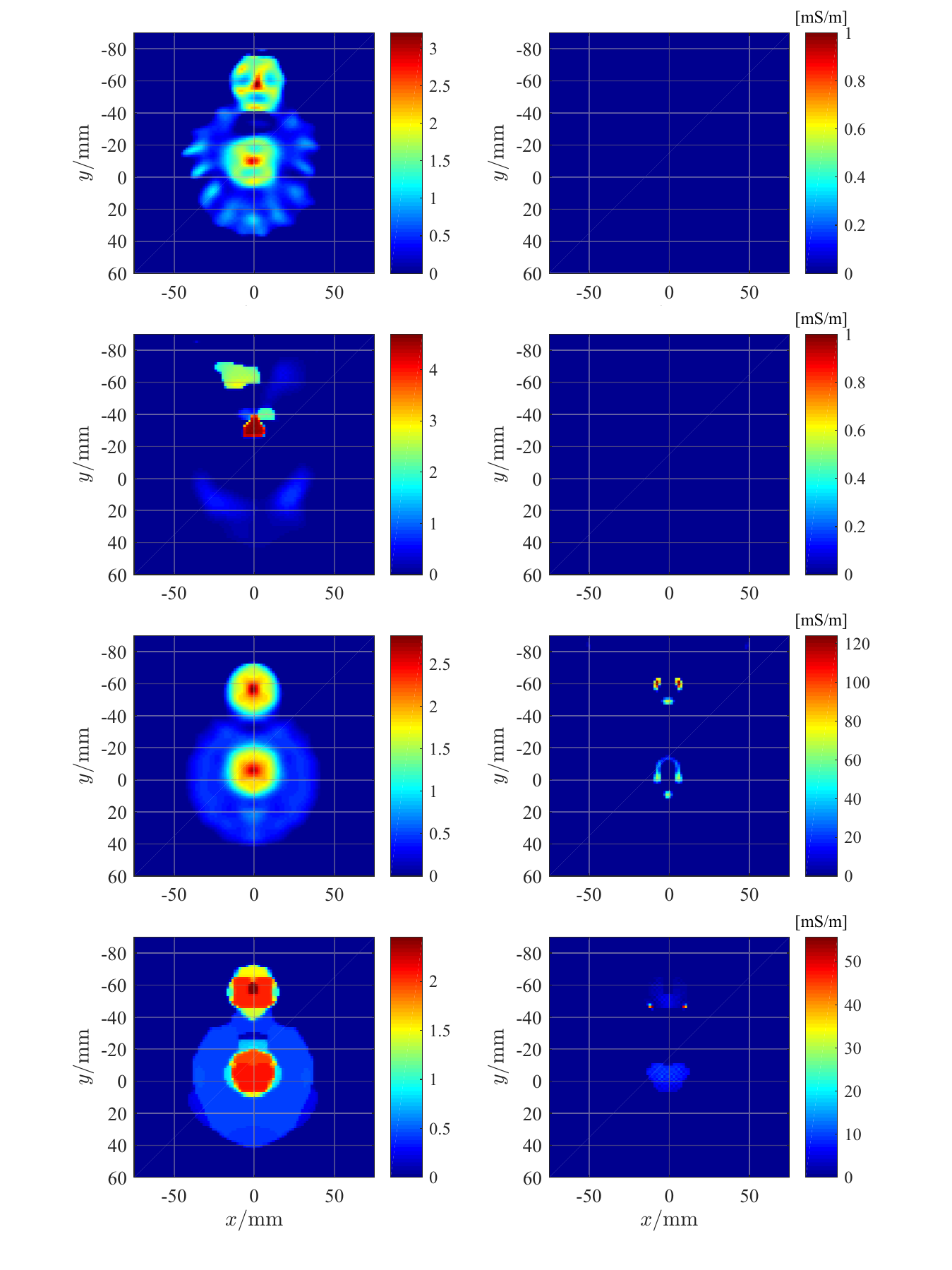}%
        \caption{Relative permittivity (left) and conductivity (right) of the inverted contrast by processing the multi-frequency dataset \textit{FoamTwinDielTE} (Line 1: MF-CC-CSI; Line 2: MF-MR-CSI) and its perfect synthetic data (Line 3: MF-CC-CSI; Line 4: MF-MR-CSI) at 7 GHz, 8 GHz, 9 GHz, and 10 GHz with 2048 iterations.}
        \label{fig:MFFoamTwinDielTEInv}
    \end{figure}

    \begin{figure}[!t]
        \centering
        \includegraphics[width=0.60\linewidth] {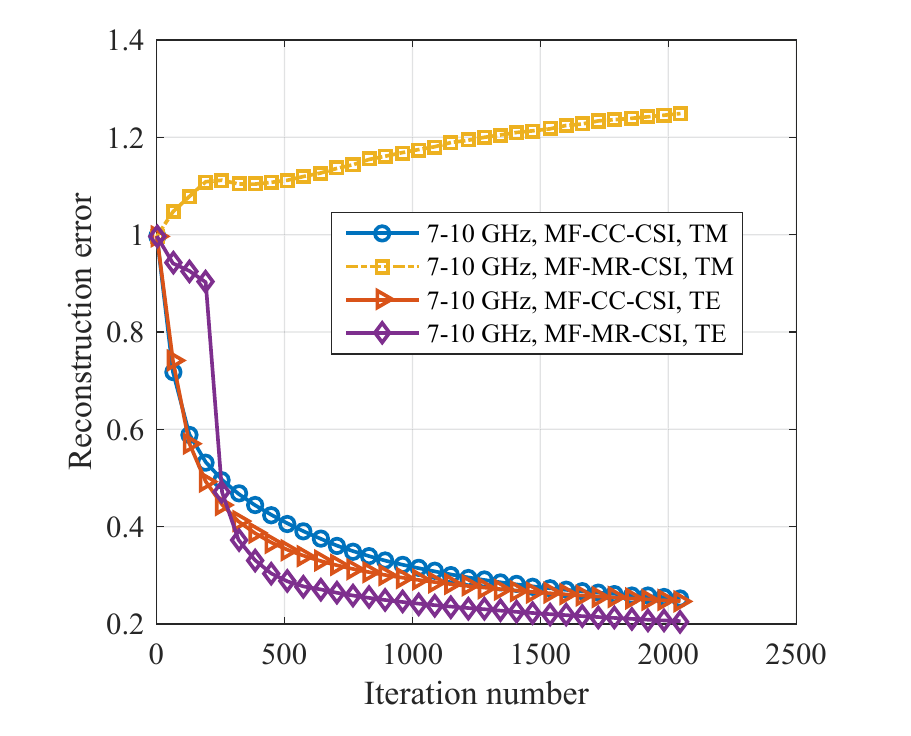}%
        \caption{Inversion error curves of MF-CC-CSI and MF-MR-CSI in terms of iteration number in processing the multi-frequency perfect synthetic data of the Fresnel dataset, \textit{FoamTwinDielTM} and \textit{FoamTwinDielTE}, at 7 GHz, 8 GHz, 9 GHz, and 10 GHz.}
        \label{fig:TwinSim}
    \end{figure}

    First, let us model the incident fields with the same approach as we did in the numerical simulation. To reduce the computational burden, we restrict the inversion domain to [$-82.5$, $82.5$] $\times$ [$-97.5$, $67.5$] mm$^2$. The inversion domain is discretized with 1.5 $\times$ 1.5 mm$^2$ grids. The multi-frequency datasets \textit{FoamTwinDielTM} and \textit{FoamTwinDielTE} at 7 GHz, 8 GHz, 9 GHz, and 10 GHz were processed by MF-CC-CSI and MF-MR-CSI, respectively. Both of them were terminated after 2048 iterations. In addition, we also did a simulation to generate the perfect data of the same targets in the same configuration. The only thing different is that the distance between antennas and the centre of the inversion domain was shortened from 1.67 m to 0.20 m. Otherwise, the scattering domain with the frequency up to 10 GHz is too huge for a standard desktop.

    Figure \ref{fig:MFFoamTwinDielTMInv} (\textit{FoamTwinDielTM}) and Figure \ref{fig:MFFoamTwinDielTEInv} (\textit{FoamTwinDielTE}) show the inverted results by processing the multi-frequency dataset (Line 1: MF-CC-CSI; Line 2: MF-MR-CSI) and the noise-free synthetic data (Line 3: MF-CC-CSI; Line 4: MF-MR-CSI) at 7 GHz, 8 GHz, 9 GHz, and 10 GHz with 2048 iterations. One can see that MF-CC-CSI successfully reconstructed the three cylinders in good estimation accuracy with both the experimental data and the noise-free synthetic data, while MF-MR-CSI only obtained good inversion results with the noise-free TE-polarized synthetic data. The artifacts of the inverted contrast conductivity (see the right figure in Line 3 of Figure \ref{fig:MFFoamTwinDielTMInv}) is inevitable and so far can only be explained as a reconstruction error to the best of our knowledge. The higher the frequency is, the larger such artifacts would be. This is easy to understand by noting that $\bm{\chi}_i=\Delta\varepsilon_r-\text{i}\Delta\sigma/\omega_i$, i.e., a large value of the angular frequency, $\omega_i$, tolerates a large error of the contrast conductivity, $\Delta\sigma$. The synthetic data enables us to obtain the reconstruction error curves of both methods, which is shown in Figure \ref{fig:TwinSim}. The error curves show us the inversion performance of both methods during the whole process of inversion, demonstrating again the advantage of MF-CC-CSI in more complicated scenarios. One can also see that the multiplicative regularization can improve the inversion accuracy indeed when the contrast can be reconstructed to a reasonable accuracy. The reason has been fully discussed in \cite{sun2017Cross}.

\section{Conclusion}\label{sec.conclusion}

    In this paper, a multi-frequency version of the cross-correlated contrast source inversion (CC-CSI) method is proposed. By processing the numerical data and experimental data in both transverse magnetic (TM) and transverse electric (TE) polarizations, we have demonstrated the advantage of the proposed multi-frequency CC-CSI (MF-CC-CSI) method over the multi-frequency multiplicative regularized contrast source inversion (MF-MR-CSI) method. In addition, the introduction of the cross-correlated error changed the classical design of the cost functional, which is of great significance for improving the inversion performance of other algorithms that are based on the classical cost functional design.

\section*{Acknowledgment}\label{sec.ack}

    The experimental data used in the manuscript is available via this link: \url{http://iopscience.iop.org/article/10.1088/0266-5611/21/6/S09/data}. The numerical data is available via this link: \url{https://surfdrive.surf.nl/files/index.php/s/pzy4U2i6caNauKd}. The full code package of CC-CSI is now available at \url{https://github.com/TUDsun/CC-CSI}. 

    In the end, we would like to show our appreciation to the Fresnel Institute for providing open access to the experimental data.

\appendix

\section{Derivation of the parameter, \texorpdfstring{$\alpha_{p,i,\ell}$}{Lg}}\label{sec.appendix-1-2}

    First, let us rewrite the cost function $\left.\mathcal{C}_{\text{MF-CC-CSI},\ell-1/2}\right|_{\bm{j}_{p,i} = \bm{j}_{p,i,\ell-1} + \alpha_{p,i}\bm{\nu}_{p,i,\ell}}$ as follows
    \begin{equation}
        \begin{split}
        & \left.\mathcal{C}_{\text{MF-CC-CSI},\ell-1/2}\right|_{\bm{j}_{p,i} = \bm{j}_{p,i,\ell-1} + \alpha_{p,i}\bm{\nu}_{p,i,\ell}} = \sum_{i=1}^I\eta_i^{\mathcal{S}}\sum_{p=1}^P\left\|\bm{\rho}_{p,i,\ell-1/2}-\alpha_{p,i}\bm{\Phi}_i\bm{\nu}_{p,i,\ell}\right\|^2+\\
        & \qquad \sum_{i=1}^I\eta_{i,\ell-1}^{\mathcal{D}}\sum_{p=1}^P\left\|\bm{\gamma}_{p,i,\ell-1/2}+\alpha_{p,i}\left(\bm{\chi}_{\ell-1}\mathcal{M}_{\mathcal{D}}\bm{A}_i^{-1}-\bm{\bm{I}}\right)\bm{\nu}_{p,i,\ell}\right\|^2+\\
        & \qquad \left.\sum_{i=1}^I\eta_i^{\mathcal{S}}\sum_{p=1}^P\left\|\bm{\xi}_{p,i,\ell-1/2}-\alpha_{p,i}\bm{\Phi}_i\bm{\chi}_{\ell-1}\mathcal{M}_{\mathcal{D}}\bm{A}_i^{-1}\bm{\nu}_{p,i,\ell}\right\|^2\right|_{\bm{j}_{p,i} = \bm{j}_{p,i,\ell-1} + \alpha_{p,i}\bm{\nu}_{p,i,\ell}}.
        \end{split}
    \end{equation}
    Obviously, it can be further simplified in the form of
    \begin{equation}
        \left.\mathcal{C}_{\text{MF-CC-CSI},\ell-1/2}\right|_{\bm{j}_{p,i} = \bm{j}_{p,i,\ell-1} + \alpha_{p,i}\bm{\nu}_{p,i,\ell}} = \sum_{j=0}^2\sum_{i=1}^I\left(a_{p,i,j}+b_{p,i,j}+c_{p,i,j}\right)\alpha_{p,i}^j.
    \end{equation}
    Therefore, we have
    \begin{equation}
        \begin{split}
        \alpha_{p,i,\ell} &=\ \underset{\alpha_{p,i}}{\text{arg}\ \max}\ \left\{\left.\mathcal{C}_{\text{MF-CC-CSI},\ell-1/2}\right|_{\bm{j}_{p,i} = \bm{j}_{p,i,\ell-1} + \alpha_{p,i}\bm{\nu}_{p,i,\ell}}\right\}\\
        &=-\frac12\frac{a_{p,i,1}+b_{p,i,1}+c_{p,i,1}}{a_{p,i,2}+b_{p,i,2}+c_{p,i,2}}.
        \end{split}
    \end{equation}
    Note that
    \begin{equation}\label{eq.pars_a2MF}
        a_{p,i,2} = \eta_i^{\mathcal{S}}\left\|\bm{\Phi}_i\bm{\nu}_{p,i,\ell}\right\|_{\mathcal{S}}^2,
    \end{equation}
    \begin{equation}\label{eq.pars_a1MF}
        a_{p,i,1} = \left.-2\eta_i^\mathcal{S} \Re\left\{\bm{\nu}_{p,i,\ell}^H\bm{\Phi}_i^H\bm{\rho}_{p,i,\ell-1/2}\right\}\right|_{\bm{j}_{p,i} = \bm{j}_{p,i,\ell-1}},
    \end{equation}
    \begin{equation}\label{eq.pars_b2MF}
        b_{p,i,2} = \eta_{i,\ell-1}^\mathcal{D}\|\bm{\nu}_{p,i,\ell}-\bm{\chi} \mathcal{M}_{\mathcal{D}}\bm{A}_i^{-1}\bm{\nu}_{p,i,\ell}\|_\mathcal{D}^2,
    \end{equation}
    \begin{equation}\label{eq.pars_b1MF}
        b_{p,i,1} = \left. 2\eta_{i,\ell-1}^\mathcal{D} \Re\left\{\bm{\nu}_{p,i,\ell}^H\left(\bm{\chi}\mathcal{M}_{\mathcal{D}}\bm{A}_i^{-1}-\bm{I}\right)^H\bm{\gamma}_{p,i,\ell-1/2}\right\}\right|_{\bm{j}_{p,i} = \bm{j}_{p,i,\ell-1}},
    \end{equation}
    \begin{equation}\label{eq.pars_c2MF}
        c_{p,i,2} = \eta_i^{\mathcal{S}}\left\|\bm{\Phi}_i\bm{\chi}\mathcal{M}_{\mathcal{D}}\bm{A}_i^{-1}\bm{\nu}_{p,i,\ell}\right\|_{\mathcal{S}}^2,
    \end{equation}
    \begin{equation}\label{eq.pars_c1MF}
        c_{p,i,1} = \left.-2\eta_i^\mathcal{S} \Re\left\{\bm{\nu}_{p,i,\ell}^H\left(\bm{\Phi}_i\bm{\chi}\mathcal{M}_{\mathcal{D}}\bm{A}_i^{-1}\right)^H\bm{\xi}_{p,i,\ell-1/2}\right\}\right|_{\bm{j}_{p,i} = \bm{j}_{p,i,\ell-1}},
    \end{equation}
    and
    \begin{equation}
        \begin{split}
        & \bm{g}_{p,i,\ell} = -2\eta_i^\mathcal{S}\bm{\Phi}_i^H\bm{\rho}_{p,i,\ell-1/2}+2\eta_{i,\ell-1}^\mathcal{D}\left(\bm{\chi}_{\ell-1}\mathcal{M}_{\mathcal{D}}\bm{A}_i^{-1}-\bm{I}\right)^H\bm{\gamma}_{p,i,\ell-1/2}-\\
        & \qquad \left.2\eta_i^\mathcal{S}\left(\bm{\Phi}_i\bm{\chi}_{\ell-1}\mathcal{M}_{\mathcal{D}}\bm{A}_i^{-1}\right)^H\bm{\xi}_{p,i,\ell-1/2}\right|_{\bm{j}_{p,i} = \bm{j}_{p,i,\ell-1}},
        \end{split}
    \end{equation}
    it is easy to obtain
    \begin{equation}\label{eq.alpha.MF}
        \alpha_{p,i,\ell} = -\frac{\Re\left\{\left\langle \bm{g}_{p,i,\ell},\bm{\nu}_{p,i,\ell}\right\rangle_\mathcal{D}\right\}}
                  {2\left(a_{p,i,2}+b_{p,i,2}+c_{p,i,2}\right)},
    \end{equation}
    where, $a_{p,i,2}$, $b_{p,i,2}$, and $c_{p,i,2}$ are given by Equation~\eqref{eq.pars_a2MF}, Equation~\eqref{eq.pars_b2MF}, and Equation~\eqref{eq.pars_c2MF}, respectively.

%
%
%
%
%
%
%
%
%


\bibliographystyle{ieeetr}
\bibliography{main_bib.bib}

\begin{thebibliography}{10}

\bibitem{colton2013inverse}
D.~Colton and R.~Kress, {\em Inverse acoustic and electromagnetic scattering
  theory}, vol.~93.
\newblock New York, USA: Springer, 3~ed., 2013.

\bibitem{caorsi2000ACom}
S.~Caorsi, A.~Massa, and M.~Pastorino, ``A computational technique based on a
  real-coded genetic algorithm for microwave imaging purposes,'' {\em IEEE
  Transactions on Geoscience and Remote Sensing}, vol.~38, pp.~1697--1708, Jul
  2000.

\bibitem{rocca2009evolutionary}
P.~Rocca, M.~Benedetti, M.~Donelli, D.~Franceschini, and A.~Massa,
  ``Evolutionary optimization as applied to inverse scattering problems,'' {\em
  Inverse Problems}, vol.~25, no.~12, p.~123003 (41pp), 2009.

\bibitem{rocca2011Diff}
P.~Rocca, G.~Oliveri, and A.~Massa, ``Differential evolution as applied to
  electromagnetics,'' {\em IEEE Antennas and Propagation Magazine}, vol.~53,
  pp.~38--49, Feb 2011.

\bibitem{salucci2017multifrequency}
M.~Salucci, L.~Poli, N.~Anselmi, and A.~Massa, ``Multifrequency particle swarm
  optimization for enhanced multiresolution {GPR} microwave imaging,'' {\em
  IEEE Transactions on Geoscience and Remote Sensing}, vol.~55, no.~3,
  pp.~1305--1317, 2017.

\bibitem{cakoni2011linear}
F.~Cakoni, D.~Colton, and P.~Monk, {\em The linear sampling method in inverse
  electromagnetic scattering}.
\newblock Newark, Delaware, USA: SIAM, 2011.

\bibitem{crocco2012linear}
L.~Crocco, I.~Catapano, L.~Di~Donato, and T.~Isernia, ``The linear sampling
  method as a way to quantitative inverse scattering,'' {\em IEEE Transactions
  on Antennas and Propagation}, vol.~60, no.~4, pp.~1844--1853, 2012.

\bibitem{wang1989iterative}
Y.~Wang and W.~C. Chew, ``An iterative solution of the two-dimensional
  electromagnetic inverse scattering problem,'' {\em International Journal of
  Imaging Systems and Technology}, vol.~1, no.~1, pp.~100--108, 1989.

\bibitem{kleinman1992modified}
R.~Kleinman and P.~van~den Berg, ``A modified gradient method for
  two-dimensional problems in tomography,'' {\em Journal of Computational and
  Applied Mathematics}, vol.~42, no.~1, pp.~17--35, 1992.

\bibitem{kleinman1993extended}
R.~E. Kleinman and P.~van~den Berg, ``An extended range-modified gradient
  technique for profile inversion,'' {\em Radio Science}, vol.~28, no.~05,
  pp.~877--884, 1993.

\bibitem{van1997contrast}
P.~M. van~den Berg and R.~E. Kleinman, ``A contrast source inversion method,''
  {\em Inverse problems}, vol.~13, no.~6, pp.~1607--1620, 1997.

\bibitem{di2009numerical}
F.~Di~Benedetto, C.~Estatico, J.~G. Nagy, and M.~Pastorino, ``Numerical linear
  algebra for nonlinear microwave imaging,'' {\em Electronic Transactions on
  Numerical Analysis}, vol.~33, pp.~105--125, 2009.

\bibitem{van1999extended}
P.~M. van~den Berg, A.~Van~Broekhoven, and A.~Abubakar, ``Extended contrast
  source inversion,'' {\em Inverse problems}, vol.~15, no.~5, pp.~1325--1344,
  1999.

\bibitem{bauer2009iteratively}
F.~Bauer, T.~Hohage, and A.~Munk, ``Iteratively regularized {G}auss-{N}ewton
  method for nonlinear inverse problems with random noise,'' {\em SIAM Journal
  on Numerical Analysis}, vol.~47, no.~3, pp.~1827--1846, 2009.

\bibitem{sun2017Linearied}
S.~Sun, B.~J. Kooij, and A.~Yarovoy, ``Linearized three-dimensional
  electromagnetic contrast source inversion and its applications to half-space
  configurations,'' {\em IEEE Transactions on Geoscience and Remote Sensing},
  vol.~55, pp.~3475--3487, June 2017.

\bibitem{caorsi2003new}
S.~Caorsi, M.~Donelli, D.~Franceschini, and A.~Massa, ``A new methodology based
  on an iterative multiscaling for microwave imaging,'' {\em IEEE transactions
  on microwave theory and techniques}, vol.~51, no.~4, pp.~1162--1173, 2003.

\bibitem{li2013contrast}
M.~Li, O.~Semerci, and A.~Abubakar, ``A contrast source inversion method in the
  wavelet domain,'' {\em Inverse Problems}, vol.~29, no.~2, p.~025015, 2013.

\bibitem{Gurbuz2009ACom}
A.~C. Gurbuz, J.~H. McClellan, and W.~R. Scott, ``A compressive sensing data
  acquisition and imaging method for stepped frequency gprs,'' {\em IEEE
  Transactions on Signal Processing}, vol.~57, pp.~2640--2650, July 2009.

\bibitem{sun2017ALinearModel}
S.~Sun, B.~J. Kooij, and A.~G. Yarovoy, ``A linear model for microwave imaging
  of highly conductive scatterers,'' {\em IEEE Transactions on Microwave Theory
  and Techniques}, vol.~66, no.~3, pp.~1149--1164, 2018.

\bibitem{sun2018ALinear}
S.~Sun, B.~J. Kooij, A.~Yarovoy, and T.~Jin, ``A linear method for shape
  reconstruction based on the generalized multiple measurement vectors model,''
  {\em IEEE Transactions on Antennas and Propagation}, vol.~66, no.~4,
  pp.~2016--2025, 2018.

\bibitem{oliveri2011bayesian}
G.~Oliveri, P.~Rocca, and A.~Massa, ``A {B}ayesian-compressive-sampling-based
  inversion for imaging sparse scatterers,'' {\em IEEE Transactions on
  Geoscience and Remote Sensing}, vol.~49, no.~10, pp.~3993--4006, 2011.

\bibitem{Poli2013MT}
L.~Poli, G.~Oliveri, F.~Viani, and A.~Massa, ``{MT}--{BCS}-based microwave
  imaging approach through minimum-norm current expansion,'' {\em IEEE
  Transactions on Antennas and Propagation}, vol.~61, pp.~4722--4732, Sept
  2013.

\bibitem{Ambrosanio2015ACom}
M.~Ambrosanio and V.~Pascazio, ``A compressive-sensing-based approach for the
  detection and characterization of buried objects,'' {\em IEEE Journal of
  Selected Topics in Applied Earth Observations and Remote Sensing}, vol.~8,
  pp.~3386--3395, July 2015.

\bibitem{sun2017Cross}
S.~Sun, B.~J. Kooij, T.~Jin, and A.~G. Yarovoy, ``Cross-correlated contrast
  source inversion,'' {\em IEEE Transactions on Antennas and Propagation},
  vol.~65, pp.~2592--2603, May 2017.

\bibitem{bloemenkamp2001inversion}
R.~F. Bloemenkamp, A.~Abubakar, and P.~M. van~den Berg, ``Inversion of
  experimental multi-frequency data using the contrast source inversion
  method,'' {\em Inverse problems}, vol.~17, no.~6, pp.~1611--1622, 2001.

\bibitem{geffrin2005free}
J.-M. Geffrin, P.~Sabouroux, and C.~Eyraud, ``Free space experimental
  scattering database continuation: experimental set-up and measurement
  precision,'' {\em inverse Problems}, vol.~21, no.~6, pp.~S117--S130, 2005.

\bibitem{brent1973algorithms}
R.~P. Brent, {\em Algorithms for minimization without derivatives}.
\newblock Englewood Cliffs, New Jersey, 1973.

\bibitem{Forsythe1976computer}
G.~E. Forsythe, M.~A. Malcolm, and C.~B. Moler, {\em Computer Methods for
  Mathematical Computations}.
\newblock Prentice-Hall, 1976.

\bibitem{litman1998reconstruction}
A.~Litman, D.~Lesselier, and F.~Santosa, ``Reconstruction of a two-dimensional
  binary obstacle by controlled evolution of a level-set,'' {\em Inverse
  problems}, vol.~14, no.~3, pp.~685--706, 1998.

\bibitem{van2001contrast}
P.~M. van~den Berg and A.~Abubakar, ``Contrast source inversion method: state
  of art,'' {\em Journal of Electromagnetic Waves and Applications}, vol.~15,
  no.~11, pp.~1503--1505, 2001.

\bibitem{van2003multiplicative}
P.~M. van~den Berg, A.~Abubakar, and J.~T. Fokkema, ``Multiplicative
  regularization for contrast profile inversion,'' {\em Radio Science},
  vol.~38, no.~2, pp.~1--10, 2003.

\bibitem{W.Shin2013}
W.~Shin, {\em 3{D} finite-difference frequency-domain method for plasmonics and
  nanophotonics}.
\newblock PhD thesis, Stanford University, The Department of Electrical
  Engineering, USA, 2013.

\bibitem{0266-5611-17-6-301}
K.~Belkebir and M.~Saillard, ``Special section: {T}esting inversion algorithms
  against experimental data,'' {\em Inverse Problems}, vol.~17, no.~6,
  pp.~1565--1571, 2001.

\end{thebibliography}


\end{document}